# Electronic temperature and two-electron processes in overbias plasmonic emission from tunnel junctions.

*Alberto Martín-Jiménez,[1,†] Koen Lauwaet,[1] Óscar Jover,[1,2] Daniel Granados,[1] Andrés Arnau,[3,4,5]*

*Vyacheslav M. Silkin,[3,4,6] Rodolfo Miranda[1,2] and Roberto Otero[1,2]\**

[1]IMDEA Nanoscience, 28049 Madrid, Spain

[2]Depto. de Física de la Materia Condensada and Condensed Matter Physics Center (IFIMAC), Universidad Autónoma de Madrid, 28049 Madrid, Spain

[3]Donostia International Physics Center (DIPC), 20018 San Sebastián/Donostia, Spain

[4]Depto. de Polímeros y Materiales Avanzados: Física, Química y Tecnología, Facultad de Química, Universidad del País Vasco UPV/EHU, Apartado 1072, 20080 San Sebastián/Donostia, Spain

[5]Centro de Fisica de Materiales CFM/MPC (CSIC-UPV/EHU), Paseo de Manuel Lardizabal 5, 20018 San Sebastián/Donostia, Spain

[6]IKERBASQUE, Basque Foundation for Science, 48009 Bilbao, Spain

KEYWORDS: Scanning Tunneling Electroluminescence, Electronic temperature, Plasmonic nanocavities, Quasiparticle lifetime, Ag(111) surface

ABSTRACT. The accurate determination of electronic temperatures in metallic nanostructures is essential for many technological applications, like plasmon-enhanced catalysis or lithographic nanofabrication procedures. In this Letter we demonstrate that the electronic temperature can be accurately measured by the shape of the tunnel electroluminescence emission edge in tunnel plasmonic nanocavities, which follows a universal thermal distribution with the bias voltage as the chemical potential of the photon population. A significant deviation between electronic and lattice temperatures is found below 30 K for tunnel currents larger than 15 nA. This deviation is rationalized as the result of a two-electron process in which the second electron excites plasmon modes with an energy distribution that reflects the higher temperature following the first tunneling event. These results dispel a long-standing controversy on the nature of overbias emission in tunnel junctions and adds a new method for the determination of electronic temperatures and quasiparticle dynamics.

Solid state systems out of equilibrium can transiently sustain different temperatures for electronic and vibrational degrees of freedom[1–7]. Electron-electron interactions lead to a fast (~ 1 ps) thermalization of the electronic energy distribution following excitation, while comparatively weak electron-phonon scattering only allows for full thermalization with the vibrational degrees of freedom after much longer times (≥ 100 ps)[1,3,8]. Therefore, for intermediate time periods following an external excitation, the electronic cloud can have a temperature of thousands of K, while the atomic lattice remains at a much lower temperature [1–7]. This effect has been related to the enhanced catalytic and chemical activity of metallic nanoparticles under optical excitation[9,10], and it is currently being exploited to optically increase the temperature of plasmonic nanoparticles or to improve the efficiency of metal-insulator-metal junctions as THz detectors[11].

In this respect, a metallic tunnel junction traversed by an electric current is an example of out-of-equilibrium solid nanostructure. Inelastic events during the electronic flow (both during the tunneling process and elsewhere) continuously pump energy into the electronic and vibrational degrees of freedom of the junction[12–16]. Such inelastic processes can increase the system's temperature[17] (electronic, vibrational or both), but they can also create quasiparticle excitations such as localized surface plasmons which, upon radiative relaxation, lead to photon emission[15,18–21]. In principle, the maximum attainable photon energy for such electroluminescence processes is the applied bias voltage (the so-called quantum cut-off, see Figure 1b), but a clear radiation tail can be observed at higher photon energies, known as the overbias emission[22–31]. One explanation for the existence of such emission is blackbody radiation from the hot electron gas, the temperature of which would be higher than the lattice temperature due to the inelastic processes that electrons in the current flow experience[23,27,29]. Fitting this model to the experimental data requires temperatures of several thousand K to explain the observed overbias emission, a fact that has raised much interest lately [23,27,29]. However, an alternative model was also put forward to explain overbias emission: the existence of two-electron processes in which part of the energy of a first tunnel event is transferred to a second tunnel event, which can excite more energetic plasmons and, thus, lead to the emission of overbias photons[24–26,28]. This model is supported by the quadratic dependence of the overbias intensity with the tunneling current[24], and casts shadows on the purely thermal interpretation of the overbias emission and on the temperature determination of the tunnel junction. Notice that a proper functional form for the cut-off function of the tunnel luminescence spectra in plasmonic junctions has only recently been described in the literature[32] and, thus, unambiguously distinguishing thermal and non-thermal effects has been hitherto difficult.

In this Letter, we describe a thorough experimental characterization of the overbias emission of a tunnel junction between a gold tip and a Ag(111) surface (see Figure 1a) as a function of the applied bias voltage, the tunneling current and the junction temperature. Taking into consideration the expected shape of the electronic factor [32] close to the quantum cut-off, the analysis of these data demonstrates that the overbias emission spectra follows a thermal photon energy distribution, but only referred to the excess energy above the tunnel bias, i.e. the bias voltage acts as the chemical potential for overbias photons. Electronic temperatures obtained from fitting to the mentioned distribution are found to match junction temperatures above 30 K and for currents below 10 nA but deviate significantly from them for lower temperatures and higher currents. Finally, we also find that the observed electronic temperatures scale with the average time between consecutive tunneling events, i.e., with the reciprocal of the tunnel intensity, in a similar fashion as the electron energies above the Fermi level scale with the surface state electron lifetimes. All these observations suggest a two-electron mechanism for the overbias emission process in which the first tunneling electron modifies the electronic temperature of the junction, while the photon emission from the second electron samples the electronic temperature at the time it tunnels. This mechanism, thus, finds common grounds between the two controversial models for overbias emission, and offers a unique tool to follow the thermalization and quasiparticle excitations dynamics with atomic spatial resolution and ps time resolution.

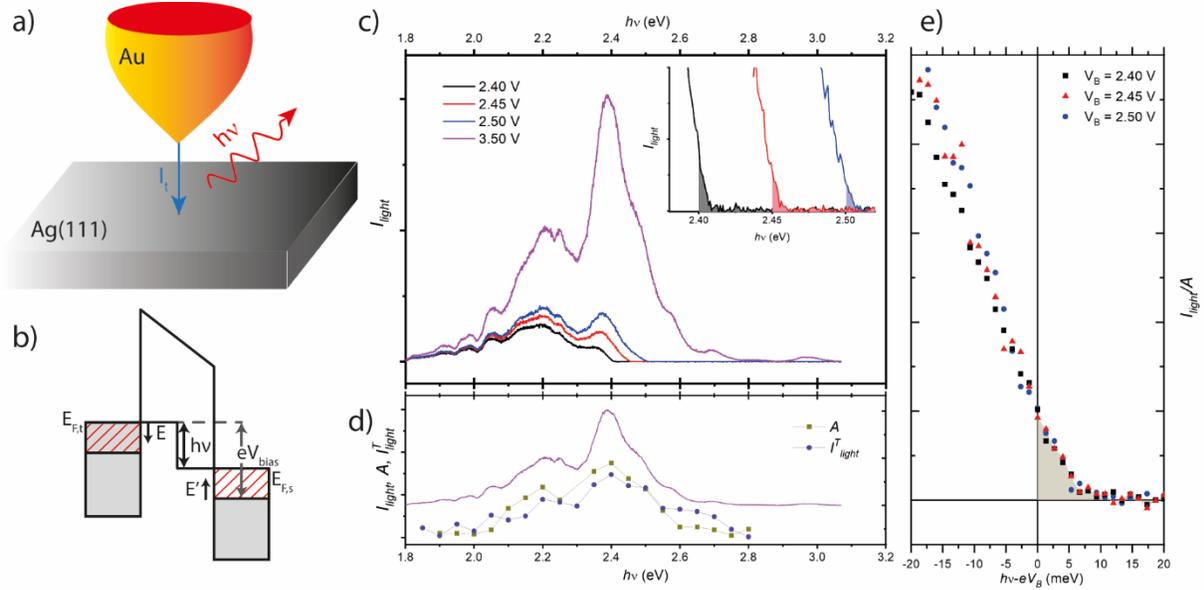

**Figure 1.** a) Schematic representation of our experiment: a tunnel current flows from a Au STM tip to a Ag(111) surface, exciting plasmons the radiative decay of which leads to photon emission. b) Level diagram showing that the width of the energy window of possible initial and final states of an inelastic tunnel process exciting a plasmon of energy $h\nu$ is $eV_{bias} - h\nu$, i.e. for small photon energies more inelastic transitions contribute to the emission. For photon energies larger than the bias voltage, inelastic processes linking occupied states in the tip and empty states in the sample become impossible. c) Tunnel electroluminescence spectra recorded at 4 K with a bias voltage of 3.5 V, where all the relevant plasmonic cavity modes can be accessed by inelastic processes, and at lower voltages (2.4-2.5 V), demonstrating the suppression of intensity at photon energies larger that the applied bias. Inset: Zoom into the emission edge. The overbias emission tail is shadowed. d) Comparison between the voltage dependence of the overbias amplitude ($A$, i.e. de light intensity at the cut-off) and total integrated emission ($I_{light}^{T}$, i.e. integrated light intensity at energies larger than the cutoff) with the fully developed spectra at 3.5

V. e) Normalization of the emission edge spectra at different voltages by their respective amplitudes makes the spectra voltage independent.

The concept of overbias emission is illustrated in Figure 1. Figure 1c displays the light intensity emitted from the tunnel junction held at 4.5 K for bias voltages of 2.4, 2.45, 2.5 and 3.5 V respectively. Since these experiments are carried out with the feedback loop closed, tip-surface distances are different for each voltage and, thus, the plasmonic modes might also change. However, according to previous investigations, the voltage range explored in this paper is narrow enough to change tip-surface distances by no more than one angstrom, and the modification of the plasmonic properties (peak intensities and positions) is negligible[32]. Notice that the energy value 3.5 eV is larger than the photon energy of all the plasmonic modes. The corresponding spectrum shows peaks at all the relevant energies, being the light intensity negligible at photon energies well below the applied bias. On the contrary, the energy values corresponding to the other three voltages sit in the middle of the plasmon-mode energies. Because the plasmonic modes of the cavity are excited by inelastic tunneling events, and the maximum energy that an electron can lose in such event is the applied bias voltage (see Figure 1b), the signal corresponding to the modes with energies higher than the bias voltage is strongly suppressed for the 2.4, 2.45 and 2.5 V cases. A careful inspection of the data, however, reveals that this suppression is not complete (see inset in Figure 1c): A small but well-defined intensity tail can be observed at photon energies above the applied bias, extending about 10 meV to the region of higher photon energies, the so-called overbias emission. For the sake of comparison, the Fermi level broadening of the tunnel spectra at 4.5 K is about 0.3 meV ($\sim 4k_BT$).

In the following, we will experimentally characterize overbias emission through two parameters, the overbias amplitude, $A$, i.e., the light intensity at the so-called quantum cut-off condition, $h\nu = eV_{bias}$, and the integrated intensity above the bias voltage, $I_{light}^T$. Figure 1d shows $A$ and $I_{light}^T$ of a tunnel junction as a function of the bias voltage, demonstrating a very structured dependence with the voltage that closely follows the shape of the fully developed light spectrum recorded with a bias voltage of 3.5 V. Indeed, we find that, by normalizing the spectral distribution of the emission edges by $A$ and plotting it versus the "excess" photon energy, $h\nu - eV_{bias}$, all the data points fall into the same curve (Figure 1e), which follows a linear dependence for relatively large negative excess energies, tends to zero for relatively large positive excess energies, and shows a transition region of about 10 meV above and below zero excess energy (the quantum cut-off region, $h\nu \sim eV_{bias}$).

We can rationalize this behavior by introducing the electronic factor that contributes to the observed tunnel electroluminescence spectra. Such spectra arise from the product of two factors, the Local Radiative Photonic Density of States of the plasmonic nanocavity at the tunnel junction, and the rate of inelastic transitions of tunneling electrons, which can be expressed as[32]

$$R_{inel}(h\nu, V_{bias}) \sim \int_{-\infty}^{+\infty} \rho_T(E - eV_{bias} + h\nu) f(E - eV_{bias} + h\nu) \rho_S(E)$$
$$\times (1 - f(E)) \mathcal{T}_{inel}(E, h\nu, V_{bias}) dE$$

(1)

where $\rho_T$ and $\rho_S$ are the tip and sample electronic density of states respectively, $\mathcal{T}_{inel}$ is the inelastic transition function and $f$ is the Fermi-Dirac function. For photon energies sufficiently close to the applied bias, the Density of States and transmission factors can be considered constant (assuming normal metal behavior, see section SI1 of Supporting Information for more details). The remaining integral involving only the Fermi-Dirac functions can readily be performed (see section SI2 of Supporting Information), and it yields:

$$R_{inel}(h\nu \sim eV_{bias}) \propto \frac{h\nu - eV_{bias}}{e^{(h\nu - eV_{bias})/k_B T_{el}} - 1} \qquad (2)$$

where $T_{el}$ is the effective temperature of the electronic cloud determining the step width of the Femi-Dirac function. Notice that, in agreement with Figure 1e, $R_{inel}$ in Equation (2) is a function of the excess energy only. Moreover, its dependence with the excess energy follows qualitatively that described in Fig. 1e for the experimental emission edges: at relatively large negative excess energies it tends to the decreasing straight line $eV_{bias} - h\nu$, while at large positive excess energies it tends to zero, with a transition region in the energy window of width $2k_B T_{el}$ around $h\nu = eV_{bias}$. The overbias emission predicted by Equation (2) results from the thermal broadening of the Femi levels. Notice that expression (2) corresponds to the average energy of an oscillatory mode of energy $h\nu - eV_{bias}$ in thermal equilibrium at temperature $T_{el}$, i.e., the bias voltage acts as the chemical potential for the creation of overbias photons, a result that should be taken into consideration when extracting temperatures from the shape of emission edges.

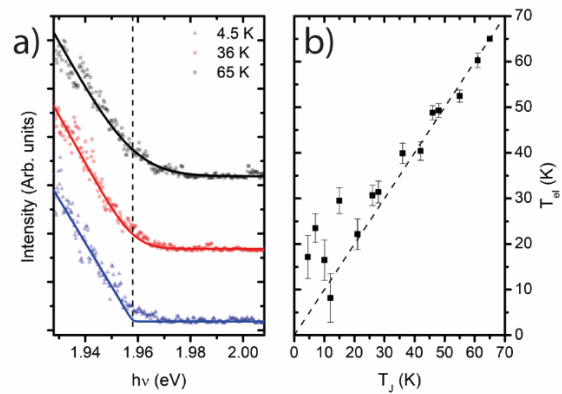

**Figure 2.** a) Experimental emission edges measured at different temperatures (black squares 65 K, red dots 36 K, blue triangles 4.5 K) with a current of 10 nA compared to the inelastic rates from Equation (2) (solid lines, same color code). b) Electronic temperatures obtained from the fitting of the experimental data sets to Equation (2) versus junction temperatures. Error bars are estimated from the standard deviation between the experimental data and the best fitted curve. The dashed line corresponds to $T_J = T_{el}$. There is a good correspondence between both measures for temperatures above 30 K, but not below.

Figure 2a shows a comparison between the expected shape of the emission edges based on Equation (2) and the experimentally observed ones for different junction temperatures, recorded with a tunnel current intensity of 10 nA. This series was recorded with the same tip and tunneling parameters by recording the tunnel electroluminescence spectra at equally spaced time intervals while slowly increasing the sample temperature in the STM. Junction temperatures ($T_J$) were measured with a diode placed on the back of the sample-holder receptacle. Here, the data have been normalized so that the slope of the edge at photon energies below the cut-off are the same for all spectra (see section SI2 in the Supporting Information for further discussion). The quantitative agreement between experimental data for a temperature of 65 K (and, somewhat less so for 35 K) and Equation (2) in Figure 2a is remarkable, supporting the interpretation of the overbias emission as arising from thermal broadening of the Fermi levels in tip and sample. The agreement between the experimentally observed emission edges and Equation (2) is further supported in the comparison between measured junction temperatures and the electronic temperatures obtained from fittings to the experimental emission edges (see Figure 2b) for junction temperatures above 30 K.

Figure 2, however, also shows significant deviations between electronic and junction temperatures at lower junction temperatures, as fitted-temperature values level off at about 17 K for final junction temperatures close to 4.5 K. As previously discussed, electronic and lattice temperatures can be significantly different in out-of-equilibrium systems, such as the tunnel junction being traversed by an electrical current. In this scenario, the rate of energy pumped into the system due to inelastic processes should increase with tunnel current intensity, and so should the electronic temperature. To test this hypothesis, we have investigated the effect of the tunneling current in the spectral shape of the emission edge for a junction held at 4.9 K (Figure 3a, the data have been normalized to the slope, as the data in Figure 2a; more information in section SI2 of the Supporting Information). Indeed, we observe that the overbias emission edges can still be fitted by Equation (2) in the range of currents between 1 and 50 nA, but the electronic temperatures and integrated overbias light intensities obtained from the fitting of Equation (2) to the experimental data do increase with increasing tunneling current (Figures 3b and c respectively). In particular, the integrated light intensity in the overbias region depends quadratically on the tunneling current, a fact that usually implies that two-electron processes are essential to describe the effect. All these results indicate that, for relatively low junction temperatures and high current intensities, the thermal broadening of the Fermi levels that determines the overbias emission is dictated by two-electron processes which raise the electronic temperature above the temperature of the lattice.

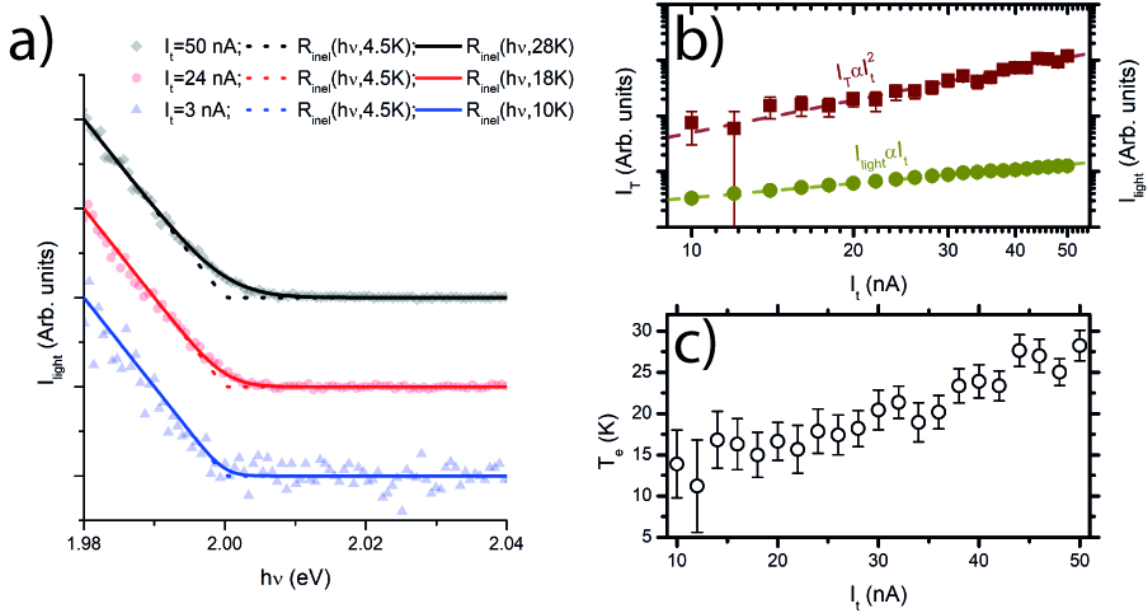

**Figure 3.** a) Emission edges as a function of the tunneling current. For comparison, the rate according to equation (2) are also plotted both using the actual temperature of the junction (4.9 K, dashed lines) or the fitted temperature (solid lines). The overbias emission becomes more intense with increasing current. b) Log-log plot of $I_{light}^{T}$ (red) and the light intensity below the quantum cut-off (precisely at 1.98 eV, green) versus the tunneling current, revealing a linear dependence on the latter case, but a quadratic dependence of the former. c) Dependence of the temperature obtained by fitting the experimental emission edges to Equation (2) for the rates as a function of the tunneling current.

This last conclusion can be related to the hot-electron hot-hole cascades previously invoked to explain overbias emission[25]. In this mechanism, which was based on experimental data obtained with very high tunneling currents of the order of µA, the tunnel injection of a first electron is followed by electron-electron and electron-hole scattering events which affects plasmon excitation by a second tunneling electron. At the relatively low tunnel intensities used here (1-50 nA), the time-lapse between the first and the second tunneling events is so large ($\tau = e/I_t$, 3-160 ps) that the energy associated to the first electron excitation has already been distributed among the rest of the electronic degrees of freedom (typically within 1 ps) but the full thermalization with the atomic lattice caused by electron-phonon scattering is still under way (typical time 100 ps). As sketched in Figure 4a, the second inelastic tunnel process might thus contribute to the overbias emission precisely because it would acquire the electronic temperature of the system at the timepoint of the second inelastic tunnel event and, indeed, this is higher than the lattice temperature. Moreover, the higher the tunnel current, the shorter the average lapse between consecutive tunnel events and therefore the thermalization time, explaining qualitatively the increase in the electronic temperature for increasing currents described in Figure 3c.

An important step to support quantitatively this mechanism is the comparison with electronic lifetimes in Ag(111). The average energy of hot electrons above the Fermi energy at a given electronic temperature can be calculated as $\langle E - E_F \rangle = \frac{\pi^2}{12\ln(2)} k_B T_{el} \approx 1.19 k_B T_{el}$ (see section SI3 in the Supporting Information). Thus, by converting from tunneling current into time span ($\tau = e/I_t$), and from electronic temperature to average energy separation from the Fermi level $\langle E - E_F \rangle$ we can reinterpret Figure 3c as a measure of the lifetimes of hot carriers as function of electron energy. Figure 4b compares such reinterpretation of the data with fist-principle calculations of the lifetime of surface state electrons in Ag(111) including only electron-phonon scattering (red squares) or including also electron-electron scattering (green circles). Inspection of Figure 4b reveals a power law dependence between average electron energy and time span between consecutive tunneling events, with an exponent (slope in the log-log representation) which is compatible with that expected from the theoretical calculations of the quasiparticle excitations in the Ag(111) surface state. The proportionality constant is smaller in the experimental data compared to the theoretical calculations by about a factor two, which we attribute to the effect of scattering with defects to limit average lifetime.

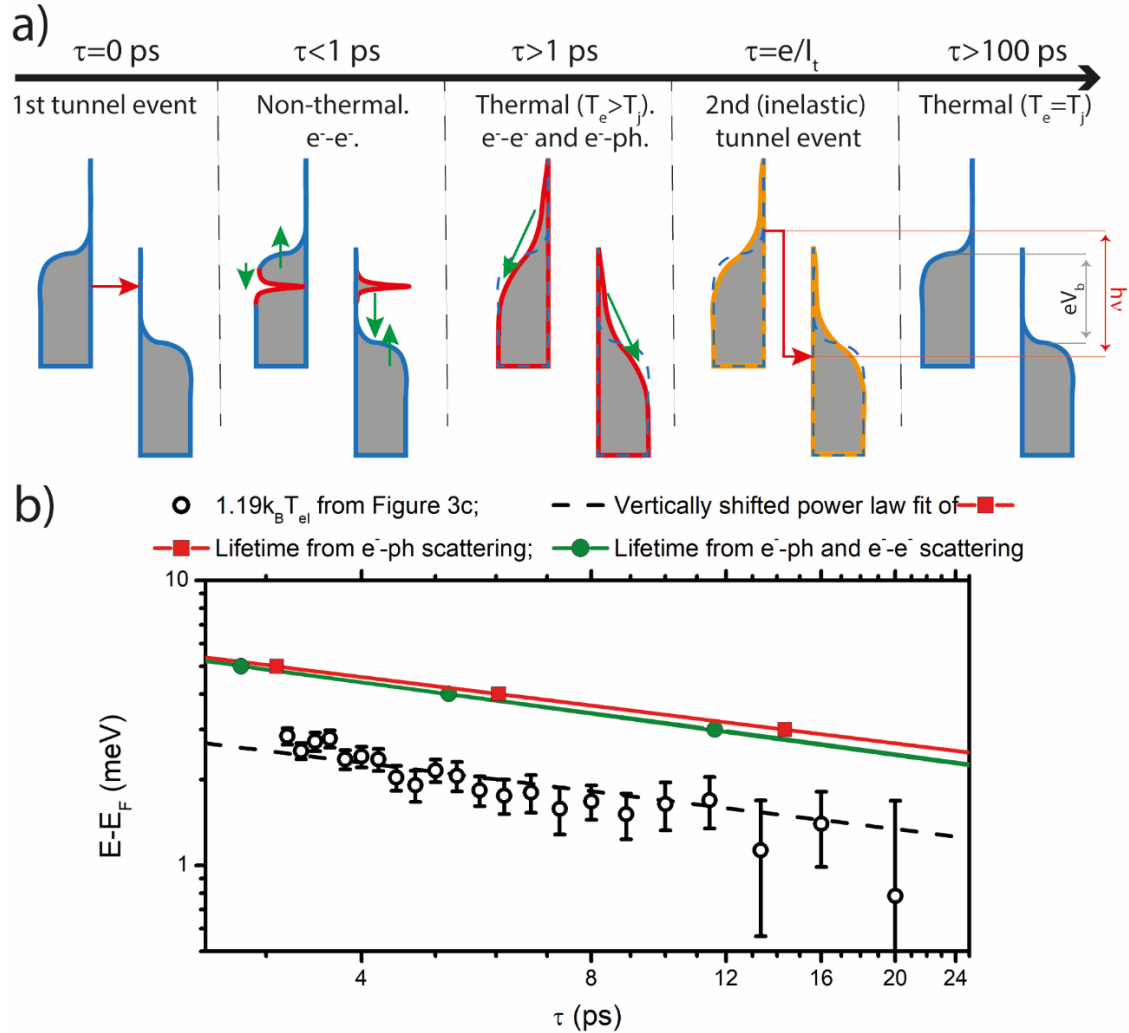

**Figure 4.** a) Proposed mechanism for the observed overbias mechanism: the first tunnel event creates hot electrons and holes, leading to a non-equilibrium occupation of the electronic states. This non-equilibrium state decays rapidly due to the e⁻-e⁻ scattering (green arrows), until the electron gas absorbs the excess energy of the hot carriers, thus increasing the electronic temperature with respect to that of the lattice. e⁻-ph scattering events now provide the tool for the electron gas to equilibrate its temperature with that of the lattice (green arrows), but full thermalization is not achieved before 100 ps. A second tunnel event taking place in this time lapse could well contribute to an overbias emission higher than the expected one in terms of lattice temperature only, because of the higher electronic temperature. b) Average energy of the electrons from the Fermi level estimated from the electronic temperatures in Figure 3c ($\langle E - E_F \rangle = 1.19 k_B T_{el}$) versus average time between tunneling events ($\tau = e/I_t$). The results are compared to calculations of the lifetime of electron quasiparticle in the Ag(111) surface state as a function of their energy, including e⁻-e⁻ and e⁻-ph scattering processes.

The agreement between the overbias analysis data and the electronic lifetime calculations strongly supports the view that overbias emission, in the range of tunneling parameters explored here, is caused by Fermi-level broadening of the emission cut-off energy due to a finite non-equilibrium electronic temperature. After the injection of a first tunneling electron, and at the moment of a second inelastic tunnel event, the value of the electronic temperature can be higher than the lattice temperature due to the incomplete thermalization of the electronic cloud. Notice that it can also be considered as a pump-probe mechanism, in which the first tunnel event creates an excitation of the system, while the second tunnel event probes the fate of the excitation after some definite time fixed by the tunnel current intensity. From that point of view, we also provide STM with a time resolution between the ps and the ns. While innovative approaches are currently lowering the time resolution of STM below the ps by applying THz and PHz pulses to the junction[33,34], and the application of electric pulses directly at the junction allow for the exploration of the ns time-scale[35,36], the time scale achievable through this technique has proven elusive to local measurement techniques.

To summarize, by means of a thorough experimental characterization and analysis of the overbias emission dependence on tunneling parameters, we have found the general shape of the emission edge spectrum and its dependence on temperature. This shape turns out to be compatible with thermal emission of plasmonic modes only when the bias voltage is considered as the chemical potential of the radiated photons. Our results show that for junction temperatures above 30 K and tunnel currents of the order of tens of nA, thermal broadening effects are sufficient to explain the overbias emission. For lower junction temperatures and higher tunneling currents, however, a significant difference is found between the electronic temperatures determined from the overbias intensities and nominal junction temperatures. All these results support a two-electron mechanism in which the first tunneling electron creates an excitation in the system, while the second one tests the thermalization of the system by the time it tunnels. This picture establishes connections between the two contradictory models previously published in the literature to explain overbias emission and opens the possibility to study the thermalization of nanoscale systems following injection of individual hot carriers with ps resolution, a method that can find applications for nanoscale metrology and for the optimization of photoactivated processes in nanostructures.

ASSOCIATED CONTENT

**Supporting Information**.

The following files are available free of charge.

Relation between the inelastic tunnel rate and the tunnel current for large and small voltages.

Functional form of the emission cut-off edge and normalization schemes. Average energy of hot electrons at finite electronic temperatures. Lifetime calculation method. (PDF)

AUTHOR INFORMATION

**Corresponding Author**

E-mail: roberto.otero@uam.es

**Present Addresses**

†Max Planck Institute for Solid State Research, Heisenbergstr. 1, 70569 Stuttgart, Germany.

**Author Contributions**

The experiments were designed by KL, AM and RO. The design and optimization of the experimental setup was carried out by AM, KL and DG. Experimental data collection was performed by AM, KL and OJ. Lifetime calculations were developed by AA and VMS. RM and RO coordinated all the work, analyzed the data, and took a leading role in the writing of the manuscript. The manuscript was written through contributions of all authors. All authors have given approval to the final version of the manuscript.

ACKNOWLEDGMENT

RM and RO acknowledge financial support from the Spanish Ministry for Economy and Competitiveness (grants PGC2018-098613-B-C21, PGC2018-096047-B-I00), the regional government of Comunidad de Madrid (grant S2018/NMT-4321), Universidad Autónoma de Madrid (UAM/48) and IMDEA Nanoscience. Both IMDEA Nanoscience and IFIMAC acknowledge support from the Severo Ochoa and Maria de Maeztu Programmes for Centres and Units of Excellence in R&D (MINECO, Grants SEV-2016-0686 and CEX2018-000805-M). AA and VMS acknowledge support from the Spanish Ministry of Science and Innovation (Grants Nos. PID2019-103910GB-I00 and PID2019-105488GB-I00, respectively) and from the Projects of the Basque Government for consolidated groups of the Basque University, through the Department of Universities (Grant Nos. IT-1246-19 and IT-1164-19, respectively).

# Electronic temperature and two-electron processes in overbias plasmonic emission from tunnel junctions.

Alberto Martín-Jiménez,[1,†] Koen Lauwaet,[1] Óscar Jover,[1,2] Daniel Granados,[1] Andrés Arnau,[3,4,5] Vyacheslav M. Silkin,[3,4,6] Rodolfo Miranda[1,2] and Roberto Otero[1,2]*

[1]IMDEA Nanoscience, 28049 Madrid, Spain

[2]Depto. de Física de la Materia Condensada and Condensed Matter Physics Center (IFIMAC), Universidad Autónoma de Madrid, 28049 Madrid, Spain

[3]Donostia International Physics Center (DIPC), 20018 San Sebastián/Donostia, Spain

[4]Depto. de Polímeros y Materiales Avanzados: Física, Química y Tecnología, Facultad de Química, Universidad del País Vasco UPV/EHU, Apartado 1072, 20080 San Sebastián/Donostia, Spain

[5]Centro de Fisica de Materiales CFM/MPC (CSIC-UPV/EHU), Paseo de Manuel Lardizabal 5, 20018 San Sebastián/Donostia, Spain

[6]IKERBASQUE, Basque Foundation for Science, 48009 Bilbao, Spain

**SI1 - Relation between the inelastic tunnel rate and the tunnel current for large and small voltages.**

For negative excess energies much larger in absolute value than $k_B T_{el}$, the expression in Equation (1) resembles that of the tunneling current at voltage $V_b - h\nu/e$. Because the tunnel current at zero bias is zero, one might naively expect that the inelastic rate and, therefore, the luminescence intensity at the cut-off should also be null. This argument, however, does not hold true because, sufficiently close to the cut-off, i.e. when the excess energies are of the order of $k_B T_{el}$, Equation (1) does not correspond to the expression for the tunneling current any longer. In this range, the correct expression for the tunnel current is

$$I_t(V_b) = \int_{-\infty}^{+\infty}[f(E-eV_b) - f(E)]\rho_T(E-eV_b)\rho_S(E)\mathcal{T}(E,V_b)dE \ . \tag{S1}$$

Equation (S1) correctly describes the null tunnel current for zero applied bias and the sense of current flowing from the electrode with the larger to bias to that with the lower bias (because $f(E-eV_b) > f(E)$ for $eV_b > 0$). The quantity in square brackets, however, can de decomposed as

$$f(E-eV_b) - f(E) = f(E-eV_b)[1-f(E)] - f(E)[1-f(E-eV_b)] \ . \tag{S2}$$

Upon substitution of Equation (S2) in Equation (S1), we can describe the tunnel current as the sum of two terms, the first describing the flow of current in the sense of the applied voltage, and a second term describing a current flowing against the voltage gradient. The second one describing the inverse current is caused by the finite temperature broadening of the Fermi levels, which enables a small but finite amount of electrons above the Fermi level of the low voltage electrode and a similarly small but finite amount of holes below the Fermi level of the high-voltage electrode. For bias voltages much larger than the thermal broadening of the Fermi levels, such term is negligible, and the tunnel current resulting from substituting Equation (S2) into Equation (S1) and neglecting the second term is completely analogous to Equation (1). This observation justifies the use of $I_t\left(V_b - \frac{h\nu}{e}\right)$ as a gauge for $R_{inel}(h\nu, V_B)$ when $|eV_b| \gg k_B T_{el}$. In the opposite limiting case, however, when $|eV_b| \sim k_B T_{el}$, the second term of Equation (S2) can no longer be neglected, leading to a completely different behavior of $R_{inel}(h\nu, V_B)$ and $I_t\left(V_b - \frac{h\nu}{e}\right)$.

The reason why a back-flowing term like the second one in Equation (S2) does not contribute to the inelastic rate factor is that the reverse process of the inelastic excitation during tunneling is the absorption of a plasmon excitation to promote an electron against the voltage gradient. Since the experiments are carried out in the dark and the average time between consecutive tunneling electrons is much larger than the average lifetime of plasmon excitation, the plasmon population before each tunnel event can be neglected and, thus, plasmon absorption-assisted back-tunneling is likewise negligible.

**SI2 – Functional form of the emission cut-off edge, normalization schemes and fitting procedures.**

Tunnel electroluminescence spectra arise from the contribution of two factors, the photonic density of radiative states, $D_r(h\nu)$, and the inelastic rate factor, $R_{inel}(h\nu, V_b)$, which describes the availability of initial and final electronic states for the inelastic tunnel process.

$$I_{light}(h\nu, V_b) = D_r(h\nu)R_{inel}(h\nu, V_b) \,. \tag{S3}$$

For small excess energies, $\varepsilon = h\nu - eV_b$ (of the order of $k_B T_{el}$), the Fermi-Dirac factor in the integrand of Equation (1) will be non-zero only for energies in a narrow window of width $\sim k_B T_{el}$ immediately above or below the Fermi level depending on the sign of the voltage. For the electronic temperatures considered here (up to 70 K), the width of this window is below 6 meV. The electronic structure of both Ag and Au are rather smooth in this energy window, much smoother than the Fermi occupancy factors. The inelastic transmission factor, likewise, will change significantly only for energies close enough to the work function. We can thus consider the electronic densities of states of tip and sample, as well as the inelastic transmission factor, constant for the integration of Equation (1), and the remaining factors can be readily and substituted into Equation (S3) to obtain the shape of the emission cut-off edge.

$$I_{light}(h\nu \sim eV_b, V_b) = KD_r(h\nu) \frac{(h\nu - eV_b)}{\left(\exp\left(\frac{h\nu - eV_b}{k_B T_{el}}\right) - 1\right)}. \tag{S4}$$

In this Equation, $K = \rho_t(0)\rho_s(0)\mathcal{T}_{inel}(0,0,0)$ is a constant which does not depend on either bias voltage nor photon energy.

Finally, we write the photonic density of radiative states as a function of the excess energy by expanding $D_r(h\nu)$ as a Taylor series around $h\nu = eV_b$, and substitute in Equation (S4) keeping only the leading term in excess energies. This approximation is justified by the fact that we are investigating the emission edges in a narrow energy window of ±20 meV, while the width of the plasmonic resonances in $D_r(h\nu)$ are at least one order of magnitude larger. This manipulation, finally, leads to the expression:

$$I_{light}(h\nu \sim eV_b, V_b) = KD_r(h\nu = eV_b) \frac{(h\nu - eV_b)}{\left(\exp\left(\frac{h\nu - eV_b}{k_B T_{el}}\right) - 1\right)}. \tag{S5}$$

Notice that this expression is just a function of the excess energy, in agreement with our experimental data in Figure 1d. Moreover, we can obtain the amplitude of the overbias, i.e. the value of the light intensity for a photon energy that equals the applied bias, can readily be calculated from Equation (S4) by taking the limit $h\nu \to eV_b$, yielding:

$$A(V_b) = K k_B T_{el} D_r(h\nu = eV_b) \tag{S6}$$

Equation (S6) correctly describes the results of Figure 1d, that is, the similar dependency of the amplitude with the bias voltage and the light intensity with the photon energy. Normalizing light intensities to the amplitude, thus, leads to

$$\frac{I_{light}(h\nu,V_b)}{A(V_b)} = \frac{1}{k_B T_{el}} (h\nu - eV_b) \Big/ \left(\exp\left(\frac{h\nu - eV_b}{k_B T_{el}}\right) - 1\right). \qquad (S7)$$

Equation (S7) only depends on the bias voltage through the excess energy and the slope in the negative excess energy region is independent of the photon energy and the bias voltage, which fully agrees with the experimental data of Figure 1e.

On the other hand, Equation (S7) also shows that normalization to the amplitude is not useful when comparing results with different values of the electronic temperature, such as those recorded with different junction temperatures or tunnel current intensity. In these cases (described in Figure 2 and 3), it is more convenient to fit the linear part at negative excess energies with a straight line. According to Equation (S5), such slope should equal $KD_r(h\nu = eV_b)$ and, normalizing to this slope one obtains exactly the second term of Equation (2) in the main text.

All the electronic temperatures in Figures 2 and 3 have been obtained by fitting the experimentally determined emission edges to equation (2), with the normalization described above. We have found that the fitting is extremely sensitive to small errors in the determination of the bias voltage, so that by changing the bias voltage by just a few meV, the fitting value of the electronic temperature can change by more than 10 K. In order to account for possible offsets in the bias voltage, we start by fitting the point at the largest temperature in Figure 2, treating the bias voltage as a fitting parameter bounded to the region separated by 10 meV from its nominal value, which is our accuracy limit. For any other fitting, the bias voltage used is no longer a fitting parameter, but it is fixed to the value obtained in the previous step.

**SI3 – Average energy of hot electrons at finite electronic temperatures.**

The average energy above the Fermi level of hot electrons at a given electronic temperature can be computed as:

$$\langle E - E_F \rangle = \frac{\int_{E_F}^{\infty}(E-E_F)\rho_S(E)f(E)dE}{\int_{E_F}^{\infty}\rho_S(E)f(E)dE} \, . \tag{S8}$$

Due to the 2D nearly-free electron gas character of electrons in the Ag(111) surface state, the density of states can be considered constant in the integration, and the resulting integrals in the numerator and denominator of Equation (S8) can be exactly calculated, yielding $\langle E - E_F \rangle = \pi^2/12\ln(2) \, k_B T_{el}$.

**SI4 – Materials and methods**.

*Sample and tip preparation*

The experiments were performed with under Ultra-High-Vacuum (UHV) conditions (P ~ 10−11 mbar), using a Omicron Low-Temperature Scanning Tunnelling Microscope (LT-STM), operated at 4.5 K, and. Clean and atomically flat Ag(111) surfaces were prepared by repeated cycles of sputtering with 1.5 keV Ar+ ions for 10 min, followed by 10 min of thermal annealing at 500 K. The Au tips were electrochemically etched in a solution of HCl (37%) in ethanol, at equal parts, and cleaned in UHV by sputtering with 1.5 keV Ar+ ions for 50 min. Our light detection set-up is formed by three lenses, one in

UHV and two in air, three mirrors and an optical spectrometer (Andor Shamrock 500) equipped with a Peltier cooled Charge-Coupled-Device (Newton EMCCD). The presented spectra are not corrected by the efficiency of the set-up.

*Lifetime calculations*

Inelastic scattering rates of the Shokley s − p surface state residing at the Ag(111) surface in a wide energy gap around the centre of the surface Brillouin zone is obtained by means of a self-consistent many-body calculation, based on the electronic self-energy. Within this approach, the inelastic linewidth (or inverse lifetime) of a quasiparticle (electron or hole) in the initial state $\phi_i(\mathbf{r})$ with energy $E_i$ is given by the projection of the imaginary part of the electron self-energy, $\Sigma$, onto this initial state [1]

$$\tau^{-1} = -2 \int d\mathbf{r} \int d\mathbf{r}' \phi_i^*(\mathbf{r}) \text{Im}[\Sigma(\mathbf{r}, \mathbf{r}', E_i)] \phi_i(\mathbf{r}') \ . \tag{S9}$$

The self-energy is computed in a so-called $GW$ approximation [2], which represents the first term in a self-energy $\Sigma$ expansion in terms of the Coulomb screened interaction, $W(\mathbf{r}, \mathbf{r}', \omega)$. Usually, in practical calculations the noninteracting Green function, $G_0$, is used instead of the exact many-body Green function $G$. This was done in present work as well whereas $W$ is obtained at the random-phase approximation (RPA) level.

The screened Coulomb interaction, $W(\mathbf{r}, \mathbf{r}', \omega)$, is related to the creation of electron-hole pairs in the system, which is the process described here through which surface states decay

$$W(\mathbf{r}, \mathbf{r}', \omega) = v(\mathbf{r}, \mathbf{r}') + \int d\mathbf{r}_1 \int d\mathbf{r}_2 v(\mathbf{r}, \mathbf{r}_1) \chi(\mathbf{r}_1, \mathbf{r}_2, \omega) v(\mathbf{r}_2, \mathbf{r}') \ , \tag{S10}$$

where $v(\mathbf{r}, \mathbf{r}')$ is the bare Coulomb interaction and $\chi(\mathbf{r}_1, \mathbf{r}_2, \omega)$ is the response function of the interacting electron system. The response function is obtained within the RPA [1,3], solving an integral equation

$$\chi(\mathbf{r}, \mathbf{r}', \omega) = \chi^0(\mathbf{r}, \mathbf{r}', \omega) + \int d\mathbf{r}_1 \int d\mathbf{r}_2 \chi^0(\mathbf{r}, \mathbf{r}_1, \omega) \frac{1}{|\mathbf{r}_1 - \mathbf{r}_2|} \chi(\mathbf{r}_2, \mathbf{r}', \omega), \quad \text{(S11)}$$

where $\chi^0(\mathbf{r}, \mathbf{r}', \omega)$ is the density-response function of noninteracting electrons [4].

After some mathematics the inverse lifetime can be expressed as the sum over all the possible final states (characterized by wave function $\phi_f(\mathbf{r})$ and energy $E_f$) of the projection over initial and final states of the imaginary part of the screened Coulomb interaction

$$\tau^{-1} = 2 \sum_f \int d\mathbf{r} \int d\mathbf{r}' \phi_i^*(\mathbf{r}) \phi_f^*(r') \text{Im}\big[-W(r, r', |E_i - E_f|)\big] \phi_i(r') \phi_f(r). \quad \text{(S12)}$$

Here we assume that the valence charge-density and the one-electron potential are constant in the $(x, y)$ directions parallel to the surface. All the variations occur in the $z$ direction perpendicular to the surface only. Then wave functions can be taken of the following form

$$\phi_n(r) = \frac{e^{i\mathbf{k}_\parallel \mathbf{r}_\parallel}}{\sqrt{A}} \psi_n(z), \quad \text{(S13)}$$

where $A$ is the normalization area of the surface and $\mathbf{k}_\parallel$ is the two-dimensional momentum parallel to the surface. For the energy dispersion we assume the next in-plane dependence

$$E_n = E_n^0 + \frac{\hbar^2 k_\parallel^2}{2 m_n}, \quad \text{(S14)}$$

where the electron effective mass $m_n$ in each subband $n$ is fitted to the photoemission experiment. The wave functions $\psi_n$ and energies $E_n^0$ of the one-particle electron states (both the bulk-like and surface ones) at the Ag(111) surface are calculated employing a single-slab geometry. The slab containing 51 atomic layers is used. The s–p one-electron potential is taken in the form proposed in [5]. In the evaluation of density-response function we included the effects due to valence 4d electrons as described in [6].

The resulting expression for the scattering rate takes the next form

$$\tau^{-1} = 2\sum_f \int \frac{d\mathbf{q}_\parallel}{(2\pi)^2} \int dz \int dz' \psi_i^*(z)\psi_f^*(z')\text{Im}[-W(z,z',\mathbf{q}_\parallel,|E_i - E_f|)]\psi_i(z')\psi_f(z)$$

(S15)

with $W(z, z', \mathbf{q}_\parallel, |E_i - E_f|)$ being the two-dimensional Fourier transform of the screened Coulomb interaction. In practical calculations, we transform all the quantities defined in the real space into the matrices performing Fourier transformation. Further details can be found in [7].